\documentclass[twocolumn,superscriptaddress,prb,amsmath,amssymb]{revtex4}
\usepackage{graphicx}% Include figure files
\usepackage{dcolumn}% Align table columns on decimal point
\usepackage{bm}% bold math

\begin{document}

%\preprint{APS/123-QED}

\title{Torque magnetometry studies of metamagnetic transitions in
single-crystal \\ HoNi$_{2}$B$_{2}$C and ErNi$_{2}$B$_{2}$C at
$T\approx$~1.9~K
}% Force line breaks with \\

\author{D. G. Naugle}
\email{naugle@physics.tamu.edu} \affiliation{Physics Department,
Texas A\&M University, TX 77843, USA}
 %\altaffiliation[Also at ]{Texas 77843-4242, USA}

\author{B. I. Belevtsev}
\email{belevtsev@ilt.kharkov.ua} \affiliation{B. Verkin Institute
for Low Temperature Physics and Engineering, National Academy of
Sciences, pr. Lenina 47, Kharkov 61103, Ukraine}

\author{K. D. D. Rathnayaka}
\affiliation{Physics Department, Texas A\&M University, TX 77843,
USA}

\author{S. A. Adegbenro}
\affiliation{Physics Department, Texas A\&M University, TX 77843,
USA}

\author{P. C. Canfield}
\affiliation{Ames Laboratory and Iowa State University, Ames, IA
50011, USA}

\author{S.-I. Lee}
\affiliation{National Creative Research Center for
Superconductivity and Department of Physics, Pohang University of
Science and Technology, Pohang 790-784, Republic of Korea}

\begin{abstract}
The metamagnetic transitions in single-crystal rare-earth nickel
borocarbide HoNi$_{2}$B$_{2}$C and ErNi$_{2}$B$_{2}$C have been
studied at 1.9 K with a Quantum Design torque magnetometer.  The
critical fields of the transitions depend crucially on the angle
between applied field and the easy axis [110] for HoNi$_2$B$_2$C
and [100] for ErNi$_2$B$_2$C. Torque measurements have been made
while changing angular direction of the magnetic field (parallel
to basal tetragonal $ab$-planes) in a wide angular range (more
than two quadrants). The results are used not only to check and
refine the angular diagram for metamagnetic transitions in these
compounnds, but also to find new features of the metamagnetic
states. Among new results for the Ho borocarbide are the influence
of a multidomain antiferromagnetic state, and ``frustrated''
behavior of the magnetic system for field directions close to the
hard axis [100]. Torque measurements of the Er borocarbide clearly
show that the sequence of metamagnetic transitions with increasing
field (and the corresponding number of metamagnetic states)
depends on the angular direction of the magnetic field relative to
the easy axis.
\end{abstract}

%\pacs{Valid PACS appear here}
\maketitle

The rare-earth nickel borocarbides of the type RNi$_{2}$B$_{2}$C
(where R is a rare-earth element) have attracted considerable
interest in the last decade because of their unique
superconducting and/or magnetic properties. In this article, a
torque magnetometry study of metamagnetic transitions at low
temperature ($T\approx 1.9$~K) in single-crystal borocarbides
with R = Ho and Er is presented. Magnetic states in these
magnetic superconductors  are determined by magnetic moments of R
ions which lay in the $ab$-planes, aligning along easy axes,
which are [110] for Ho and [100] for Er,
respectively resulting in a high magnetic anisotropy for these
compounds.
\par
In zero field, HoNi$_{2}$B$_{2}$C is a superconductor (below
$T_c\approx 8.7$~K) and antiferromagnetic (AFM) (below the N\'{e}el
temperature, $T_N \approx 5.5$~K). For $T\lesssim 4$~K, with
increasing magnetic field $H$ (perpendicular to the tetragonal
$c$-axis), the sequence of transitions from antiferromagnetic
($\uparrow\downarrow$) to ferrimagnetic
($\uparrow\uparrow\downarrow$), non-collinear
($\uparrow\uparrow\rightarrow$) and ferromagnetic-like
($\uparrow\uparrow$) states takes place at critical fields
$H_{m1}$, $H_{m2}$ and $H_{m3}$, respectively \cite{canf}. Neutron
diffraction studies show that these magnetic states are $c$-axis
modulated except for the non-collinear phase which is $a$-axis
modulated \cite{camp,detl}.
\par
In ErNi$_{2}$B$_{2}$C, superconductivity and antiferromagnetism
coexists as well ($T_c\approx 11$~K and $T_N\approx 6$~K). Below
$T_N$ the magnetic phases are spin-density wave (SDW) states with
magnetic wave vector $\mathbf{Q}= f \mathbf{a^{*}}$ (or
$\mathbf{b^{*}}$, where $\mathbf{a^{*}}$ and $\mathbf{b^{*}}$ are
reciprocal lattice vectors) \cite{jensen}. It was found that
$f\approx 0.55$ for the AFM state. Below $T_{WF}\approx 2.5$~K a
transition to a weak-ferromagnetic (or, actually, ferrimagnetic)
state takes place, in which a ferromagnetic moment (about 0.33
$\mu_B$ per Er ion) appears. With increasing field (applied in the
$ab$ plane) several metamagnetic transitions occur in this
compound. Generally, the ferromagnetic component increases at each
transition reaching the maximum value (about 8 $\mu_B$/Er) at
the final transition to a saturated paramagnetic (or
ferromagnetic-like) state at $H\gtrsim 2$~T. The resulting
metamagnetic states (except the ferromagnetic-like state) remain SDW,
 only the scalar $f$ of the wave vector $\mathbf{Q}= f
\mathbf{a^{*}}$ changes slightly, but quite distinctly, at these
transitions. The only longitudinal magnetization study of these
transitions \cite{canf2} revealed three metamagnetic transitions
(and correspondingly three SDW magnetic structures) with
increasing field in this compound, independent of the angle between
$H$ and the easy axis. Subsequent neutron diffraction studies
\cite{jensen} have shown, however, that for $H\|$[010] four SDW
magnetic structures can be distinguished for different field
ranges; whereas, only three SDW phases are seen for $H\|$[110]
([010] and [110] are the easy and hard axes, respectively).
\par
The critical fields of metamagnetic transitions depend strongly on
the angle $\theta$ between $H$ and the nearest easy axis (or on the
angle $\phi$ between $H$ and the nearest hard axis) for both
borocarbides \cite{canf,canf2}. In this study, a PPMS Model 550
Torque Magnetometer (Quantum Design) was used to study this
dependence. It measures the torque $\vec{\tau} =
\mathbf{M}\times\mathbf{H}$, so that $\tau = MH \sin(\beta)$,
where $\beta$ is the angle between the external magnetic field and
the magnetization.  Small (0.05-0.15 mg) single-crystal
rectangular plates of HoNi$_2$B$_2$C ($0.4\times 0.32\times 0.26$
mm$^{3}$) and ErNi$_2$B$_2$C ($0.37\times 0.32\times 0.25$
mm$^{3}$) were cut and polished for this study. The torque
measurements were made under changes of magnetic field for
different constant angles, or changes of angular direction of the
magnetic field for different constant magnetic fields.

\begin{figure}[t]
\includegraphics[width=0.7\linewidth]{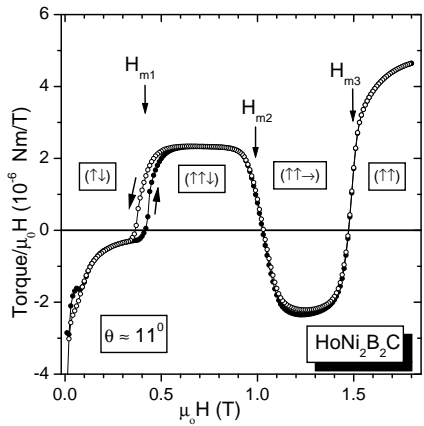}% Here is how to import EPS art
\caption{Field dependence $\tau(H)/H$ in a HoNi$_2$B$_2$C sample,
recorded for increasing and decreasing magnetic field (at $\theta
\approx 11^{\circ}$). Positions of the critical transition fields
$H_{m1}$, $H_{m2}$ and $H_{m3}$ are shown by arrows. The symbols
($\uparrow\downarrow$), ($\uparrow\uparrow\downarrow$),
($\uparrow\uparrow\rightarrow$) and ($\uparrow\uparrow$) show
areas of different magnetic phases.}
\end{figure}

\par
An example of the magnetic-field dependence for HoNi$_2$B$_2$C is
shown in Fig. 1, where transitions manifest themselves as sharp
changes of the torque at critical fields. The angular phase
diagram for HoNi$_2$B$_2$C based on these torque
measurements is found to correspond generally to known experiment
\cite{canf} and theoretical models \cite{kalats,amici}. But
important new features of these metamagnetic states were found and
other features made clearer as indicated below under items A, B,
and C:

\par
{\bf (A)} According to Ref. \onlinecite{canf}, for small
deviations of the magnetic field from a $\langle 110\rangle$ axis
($-6^{\circ}\leq \theta \leq 6^{\circ}$), the
($\uparrow\downarrow$)--($\uparrow\uparrow\downarrow$)--($\uparrow\uparrow$)
sequence of transitions takes place. In this sequence, the
transition to the non-collinear ($\uparrow\uparrow\rightarrow$)
phase is omitted. This is in disagreement with theory
\cite{amici}, which supposes that this sequence of transitions is
possible at $\theta = 0$ only. Analysis of the magnetic-field
dependences of the torque for different angles, including angles
close to $\theta = 0$, leads to the conclusion that the angular
range for this sequence of metamagnetic transitions is far
less ($-1^{\circ}\leq\theta\leq 1^{\circ}$) than that indicated in
Ref. \onlinecite{canf}.

\begin{figure}
\includegraphics[width=0.86\linewidth]{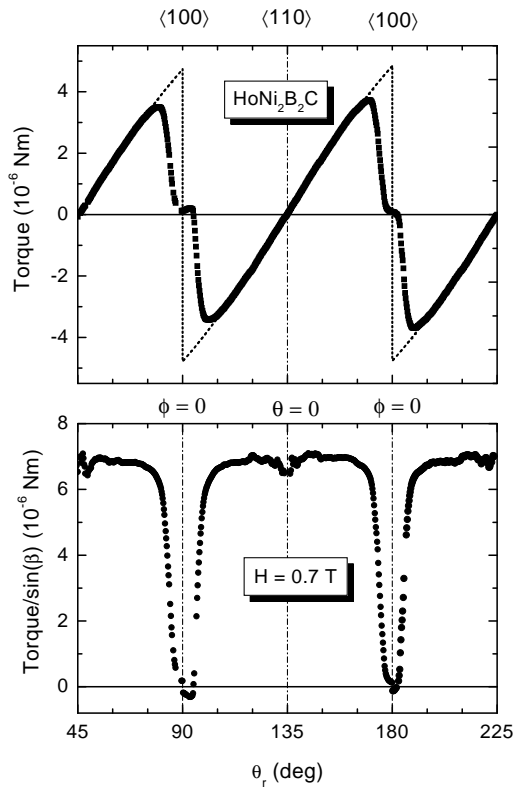}% Here is how to import EPS art
\caption{Angular dependences of the torque (upper panel) and
$\tau/\sin(\beta)$ (bottom panel) at the field $\mu{_0}H=0.7$~T in
HoNi$_2$B$_2$C. The angle $\theta_{r}$ represents the angular
position of the rotator. The positions of the easy $\langle 110
\rangle$ and hard $\langle 100 \rangle$ axes together with the
corresponding positions of $\theta=0$ and $\phi=0$, are shown. The
dashed line indicates the expected dependence of the torque if the
transition near $\phi=0$ from one orientation of the ferrimagnetic
state ($\uparrow\uparrow\downarrow$) to another
($\rightarrow\rightarrow\leftarrow$) were sharp.}
\vspace{-10pt}
\end{figure}

\par
$\bf (B)$ The magnetization of the antiferromagnetic
($\uparrow\downarrow$) phase must be equal to zero, and the same
should be expected for the torque (for $H < H_{m1}$). It is found,
however, that the magnitude $\tau(H)/H$ is non-zero below
$H_{m1}$. Moreover, the $\tau(H)/H$ curves also show an
appreciable field dependence with hysteretic behavior in the
low-field region $\mu_{0}H < 0.1$~T (Fig. 1). This type of
$\tau(H)/H$ dependence is found for the entire angular region
studied. The non-zero absolute value of $\tau(H)/H = M\sin(\beta)$
implies that $M \neq 0$ as well. This is possible for the AFM
state if a multidomain AFM structure exists. This may result from
availability of four (or at least two) equivalent easy $\langle
110\rangle$ directions in HoNi$_{2}$B$_{2}$C. Then, on cooling
below the N\'{e}el temperature, domains can easily appear. The
low-temperature orthorhombic distortions in borocarbides
\cite{krei} could facilitate this process. When multidomain (or at
least two-domain) AFM structures exist, the magnetic moments of
the domains may not be completely compensated, and the torque can
be non-zero. It should be noted that the torque hysteresis in the
AFM phase takes place in the superconducting state. In this case,
the non-zero torque and the hysteresis in the low-field range
(Fig. 1) may also be related to trapped flux generated on passing
through the critical field.
\par
{\bf (C)} The angular behavior of $H_{m2}$ for the angles close to
the hard axis $\langle 100\rangle$ ($-6^{\circ} \leq \phi \leq
6^{\circ}$) does not follow the theoretical relation $H_{m2}(\phi)
= H_{m2}(0)/\cos (\phi)$ \cite{kalats,amici}, so that the
$H_{m2}$ values in this region are smaller than predicted. Also,
in this angular range the first
($\uparrow\downarrow$)--($\uparrow\uparrow\downarrow$) and
second
($\uparrow\uparrow\downarrow$)--($\uparrow\uparrow\rightarrow$)
metamagnetic transitions almost merge together. This means that
the width of the field range, where the ferrimagnetic
($\uparrow\uparrow\downarrow$) phase exists, is reduced
substantially for small $\phi$. This ''frustrated'' behavior of
the magnetic system shows itself in the angular torque dependences
as well. In Fig. 2, the angular dependences of the torque and
$\tau/\sin(\beta)$ for $\mu{_0}H=0.7$~T is presented. At this field
only the ferrimagnetic ($\uparrow\uparrow\downarrow$) phase should
exist for any angle \cite{canf}, hence the relation $\beta=\theta$
is expected. Since $\sin(\beta)$ changes sign on crossing the
angle $\phi=0$ (due to reorientation of Ho moments relative to
the nearest easy axis), the torque should change sign as well (Fig.
2). But $\tau(\phi)$ behavior in the angular region close to $\phi=0$
does not correspond to expected behavior (shown by by the dashed
line) for the case $\beta=\theta$. It can be seen that
$\tau/\sin(\beta)$ which should be equal to the net magnetization
at finite $\beta$ goes to zero when approaching $\phi = 0$.
Outside this region, $\tau/\sin(\beta)$ is approximately constant
as expected for the ($\uparrow\uparrow\downarrow$) phase. If
$\beta \neq 0$, this would suggest that the magnetization tends
to zero when the magnetic field direction approaches $\phi = 0$.
Another possibility is that $\sin(\beta)$ goes to zero as the
field direction crosses the angle $\phi=0$. In this case the
magnetization at $\phi=0$ can be non zero, but the magnetization
should be directed along the hard axis $\langle 100\rangle$.
Longitudinal measurements of $\mathbf{M}$ support the
latter \cite{canf}. This type of the torque behavior near $\phi =
0$ was also found for higher fields where the non-collinear phase
exists. The large width of the frustration region illustrated in
Fig. 2, may be due to inhomogeneity induced by strain, defects or
demagnetization effects.
Further discussion will be presented in an extended paper.

\begin{figure}[t]
\vspace{-12pt}
\includegraphics[width=0.85\linewidth]{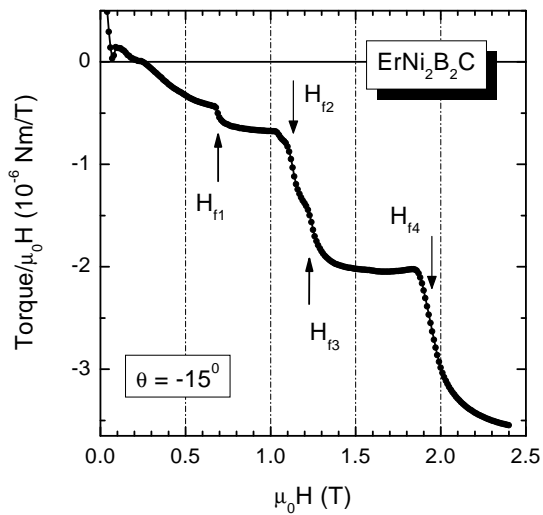}% Here is how to import EPS art
\vspace{-6pt} \caption{Field dependence $\tau(H)/H$ in
ErNi$_2$B$_2$C sample, recorded for increasing magnetic field (at
$\theta \approx -15^{\circ}$). Positions of the critical
transition fields $H_{f1}$, $H_{f2}$, $H_{f3}$ and $H_{f4}$ are
shown by arrows.
 }
\end{figure}
\par
Preliminary results for ErNi$_2$B$_2$C are shown in Figs. 3, 4.
Typical $H$ dependence of the torque for
angles not far from an easy axis is shown in Fig. 3. The torque
magnitude increases with $H$ and its behavior shows clearly four
metamagnetic transitions, indicated by arrows. The transition
fields $H_{f1}(\theta)$ and $H_{f2}(\theta)$ are found to be
proportional to $1/\cos(\theta)$ in agreement with results of
longitudinal magnetization measurements \cite{canf2}. On the other
hand, that study \cite{canf2} revealed clearly only one transition
in the range 1.1--1.4 T; whereas, the results here show
two distinct transitions (Figs. 3 , 4). For angles close
enough to the hard $\langle 110 \rangle$ axis ($-15^{\circ}\lesssim
\phi \lesssim 15^{\circ}$) only one transition is seen in this
field range (Fig. 4). The angular
range in which only one transition takes place in this field range
is thus determined. These results are consistent with the
neutron diffraction study \cite{jensen} for angles $\theta=0$ and
$\theta = 45^{\circ}$ (or $\phi=0$).   Results of the torque study below the first
transition field and for higher fields where the transitions to the
paramagnetic state take place will be considered elsewhere.

\begin{figure}
\includegraphics[width=0.85\linewidth]{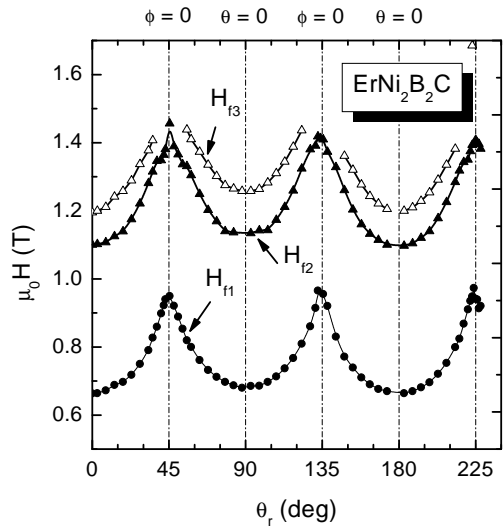}
\vspace{-4pt} \caption{The angular phase diagram of metamagnetic
states in ErNi$_2$B$_2$C at $T= 1.9$~K, obtained in this study.
Results for measurements with increasing field are shown. The
angle $\theta_{r}$ is the modified angle on the sample rotator.
The angles $\theta_{r}=0^{\circ}$, $90^{\circ}$ and $180^{\circ}$
correspond to different $\langle 100 \rangle$ easy axes. The hard
axes $\langle 110 \rangle$ are shifted by $45^{\circ}$.  Positions
of the transition fields $H_{f1}$, $H_{f2}$ and $H_{f3}$ are shown
by arrows.
 }
\vspace{-6pt}
\end{figure}

\par
This research was supported by the Robert A Welch Foundation
(A-0514), NSF (DMR-0103455, DMR-0422949, DMR-0315476) and CRDF
(UPI-2566-KH-03).

\end{document}